\documentclass[twocolumn,english,prl]{revtex4}
\usepackage[T1]{fontenc}
\usepackage[latin1]{inputenc}
\usepackage{amsmath}
\usepackage{graphicx}
\usepackage{amssymb}

\makeatletter
\usepackage{graphicx}
\usepackage{amssymb}

\usepackage{babel}

\usepackage{babel}
\makeatother
\begin{document}

\title{A nearly closed ballistic billiard with random boundary transmission}

\author{Igor Rozhkov, Ganpathy Murthy}

\affiliation{Department of Physics and Astronomy, University of Kentucky, Lexington,
Kentucky 40506}

\begin{abstract}
A variety of mesoscopic systems can be represented as billiard with
a random coupling to the exterior at the boundary. Examples include
quantum dots with multiple leads, quantum corrals with different kinds
of atoms forming the boundary, and optical cavities with random surface
refractive index. We study an electronic billiard with no internal
impurities weakly coupled to the exterior by a large number of leads.
We construct a supersymmetric nonlinear $\sigma$-model by averaging
over the random coupling strengths between bound states and channels.
The resulting theory can be used to evaluate the statistical properties
of any physically measurable quantity. As an illustration, we present
results for the local density of states.
\end{abstract}
\maketitle
When the size $L$ of a two-dimensional (2D) mesoscopic structure
is reduced beyond the elastic mean free path $l$ of the electrons,
all the scattering takes place at the boundary. An electron may escape
through the boundary, get reflected back specularly, or participate
in both processes. Besides quantum dots \cite{Marcus}, which can
exhibit both diffusive and ballistic behavior \cite{FerryGodnick,WhiteBook},
there are mesoscopic billiards which are ballistic by construction.
Examples include quantum corrals (QC) \cite{QuantumCorrals}, optical
corrals \cite{OpticalCorrals}, optical resonant cavities \cite{Stone1},
and the artificial atoms proposed in Ref. \cite{ResonantMagneticVortices}.

A quantum dot has boundary losses, unless the confining potential
is chosen to be infinitely high. Such losses are usually modeled by
coupling to a number (possibly infinite) of open channels \cite{OpenSystems},
although the precise details of coupling are generally unknown. In
this paper we focus on internal ballistic dynamics of a clean circular
dot which is nearly closed, i.e., it is weakly coupled to a large
number of leads. The crucial ingredient in our model is the randomness
of the transmission coefficients. It enables us to carry out an ensemble
average using the supersymmetry method \cite{Efetovsbook}, and, as
a side-benefit, acts as a natural regularizer, helping us avoid the
technical difficulties of previous supersymmetric approaches to closed
ballistic systems \cite{WhiteBook,Supersymmetry}. The resulting theory
is a {}``surface'' $\sigma$-model, which resembles the conventional
diffusive $\sigma$-model \cite{Efetovsbook}, but has the {}``diffusion''
modes confined at the boundary of the dot. These can be associated
with classical whispering gallery trajectories, which run along the
walls of the system and are known to strongly influence transport
through mesoscopic structures \cite{WGM}.

Furthermore, as pointed out in Ref. \cite{RichterSiebert}, the ballistic
analogs of density relaxation modes in diffusive samples originate
from trajectories which remain close to each other in configuration
space. Given the almost closed nature of our dot, the whispering gallery
modes (WGM) \cite{WGM} are expected to impact the long-time characteristics
of the internal dynamics and to play a major role in quantum interference
effects. By contrast, the {}``star-like'' and {}``asterisk'' trajectories,
which approach the lead mouth at small angles to the boundary normal
and are more likely to exit the billiard, cannot contribute \cite{ScweitersAlfordDelos}
to the response functions. In other examples \cite{QuantumCorrals,ResonantMagneticVortices}
the electrons do not decay into well-defined leads, but because of
the significant similarities between the three nanostructures our
approach will apply, \emph{mutatis mutandis}, to them as well. To
be specific, the long-lived modes which play a chief role in their
internal dynamics have the same classical origin.

A quantum dot in which electrons are confined by a hard wall potential
but can escape into leads \cite{footnote2} can be described by the
non-Hermitian effective Hamiltonian\begin{equation}
H_{eff}=\mathcal{H}_{0}\mp\frac{i}{2}\sum_{n=1}^{N_{\textrm{leads}}}\widehat{B}_{n}\delta_{C_{n}},\label{effectiveHam}\end{equation}
 Here, $\mathcal{H}_{0}$ is the Hamiltonian of the closed quantum
dot, and $\delta_{C_{n}}$ is a surface $\delta$-function with uniform
support on the crossection $C_{n}$ which separates the $n$th lead
from the billiard. It is important to note that the leads have been
{}``integrated out''. In other words, one replaces an open dot with
plane wave boundary conditions in the asymptotic region and Dirichlet
conditions along the rest of the boundary, with a closed dot and simplified
boundary conditions \cite{footnote2}. The operators $\widehat{B}_{n}$
are defined through their action on an arbitrary function\[
\widehat{B}_{n}\Psi\left(\mathbf{r}\right)=\int_{C_{n}}\sum_{i}^{M_{n}^{\textrm{channels}}}\gamma_{n}^{\left(i\right)}v_{n}^{\left(i\right)}\varphi_{n}^{\left(i\right)}\left(\mathbf{r}\right)\varphi_{n}^{\left(i\right)}\left(\mathbf{r}'\right)\Psi\left(\mathbf{r}'\right)d\mathbf{r}',\]
 where $\gamma_{n}^{\left(i\right)}<1$ is a coupling coefficient,
$v_{n}^{\left(i\right)}$ is a transverse velocity, and $\varphi_{n}^{\left(i\right)}$
is a normalized eigenfunction of transverse motion for $i$th channel
in $n$th lead. 

Thus, the solution of the original open dot problem reduces to the
solution of\begin{equation}
\left(E-H_{eff}\right)G^{R,A}\left(\mathbf{r},\mathbf{r}'\right)=\delta\left(\mathbf{r}-\mathbf{r}'\right),\label{GoverningEq1}\end{equation}
 for the retarded (advanced) Green's function $G^{R,A}$, with Neumann
boundary conditions $\left(\mathbf{x}_{n}\mathbf{v}_{r}\right)\left.G^{R,A}\left(\mathbf{r},\mathbf{r}'\right)\right|_{C_{n}}=0$,
where $\mathbf{x}_{n}$ is a unit vector parallel to the waveguide
walls, and $\mathbf{v}_{r}$ is the velocity operator. It acts from
the side of the lead.

Just as in studies of chaotic scattering \cite{Scattering}, the key
tool in calculations of response functions of ballistic (or disordered)
systems with leads is the operator $\left(E-H_{eff}\right)^{-1}$.
In the framework of the supersymmetry method it can be treated in
the same way as $\left(E-H_{0}\right)^{-1}$, as was done in Ref.
\cite{footnote2}, where nonperturbative calculations were carried
out for disordered dots. Note, that we pursue the non-universal, i.e.
non-''random matrix theory'' regime.

For a ballistic dot the dynamics is governed by the operator: $\mathcal{H}_{0}=\left(\mathbf{p}-\left(e/c\right)\mathbf{A}\right)^{2}$,
where we assumed no potential and a constant magnetic field. The latter
is introduced to break time reversal symmetry, which simplifies the
application of the supersymmetry method, and makes the illustration
of our approach more transparent.

We intend to calculate averages over the Gaussian distribution of
dimensionless coupling coefficients $\gamma_{n}^{\left(i\right)}$.
The relation of these coefficients to sticking probabilities, transmission
coefficients and other commonly used parameters can be found in Ref.
\cite{Scattering} in the context of the Hamiltonian approach to chaotic
scattering. More insight on the physical meaning of these coefficients
can be gained from Refs. \cite{LossySystems}. 

Before commencing our supersymmetric derivation, we make a few simplifying
assumptions. These assumptions can all be relaxed without affecting
the physics, but are needed to simplify the technical details. We
allow only one open channel in each lead and express the coupling
coefficients as a sum of constant and stochastic parts: $\gamma_{n}=\overline{\gamma}+\widetilde{\gamma}_{n}$.
For the statistics of $\widetilde{\gamma}_{n}$ we assume that $\left\langle \widetilde{\gamma}_{n}\right\rangle =0,$
$\left\langle \widetilde{\gamma}_{n}\widetilde{\gamma}_{m}\right\rangle =x^{2}\delta_{nm}$,
and that all higher moments factorize into second moments. Next, we
{}``eliminate the leads'' \cite{footnote2}, passing to the Hamiltonian
given by Eq. (\ref{effectiveHam}). 

Upon introduction of four-component supervectors $\Psi\left(\mathbf{r}\right)^{T}=\left\{ S_{1}\left(\mathbf{r}\right),\chi_{1}\left(\mathbf{r}\right),S_{2}\left(\mathbf{r}\right),\chi_{2}\left(\mathbf{r}\right)\right\} $,
and supermatrices $L=\textrm{diag}\left\{ 1,1-1,1\right\} $, $\Lambda=\textrm{diag}\left\{ 1,1-1,-1\right\} $
(see, for example, Ref. \cite{Mirlin}), the supersymmetric generating
functional reads \cite{footnote3}\begin{equation}
\left\langle Z\left[J\right]\right\rangle _{\widetilde{\gamma}_{n}}=\int d\Psi^{*}d\Psi\textrm{e}^{-\mathcal{L}\left[\Psi\right]}\left\langle \textrm{e}^{-\mathcal{L}_{\delta}\left[\Psi\right]}\right\rangle _{\widetilde{\gamma}_{n}},\label{generatingfunctional}\end{equation}
 where $\left\langle \dots\right\rangle _{\widetilde{\gamma}_{n}}$
indicates averaging over random couplings to the leads\begin{align*}
\mathcal{L}\left[\Psi\right] & =i\int\Psi^{\dagger}\left(\mathbf{r}\right)\widehat{\mathcal{H}}L\,\Psi\left(\mathbf{r}\right)d\mathbf{r}\\
 & +\frac{\overline{\gamma}}{2}\sum_{n=1}^{N}v_{n}\int_{C_{n}}\Psi^{\dagger}\left(y_{n}\right)\varphi_{n}\left(y_{n}\right)\varphi_{n}\left(y'_{n}\right)\Lambda L\,\Psi\left(y'_{n}\right),\end{align*}
 \[
\mathcal{L}_{\delta}\left[\Psi\right]=\sum_{n=1}^{N}\frac{\widetilde{\gamma}_{n}v_{n}}{2}\int_{C_{n}}\Psi^{\dagger}\left(y_{n}\right)\varphi_{n}\left(y_{n}\right)\varphi_{n}\left(y'_{n}\right)\Lambda L\,\Psi\left(y'_{n}\right).\]
 with $\widehat{\mathcal{H}}=\widehat{\mathcal{H}}_{0}I_{4}+i\epsilon\Lambda$,
($\epsilon$ is infinitesimally small) $\widehat{\mathcal{H}}_{0}=-\nabla^{2}/2m-E$,
$\varphi_{n}\left(y\right)=\sqrt{2/d_{n}}\sin\left(\pi y/d_{n}\right)$
(for hard-wall lead of width $d_{n}$). Here $\int_{C_{n}}$ stands
for a double integration over $y_{n}$ and $y'_{n}$, the transverse
coordinates along the crossection $C_{n}$ (perpendicular to the walls
of a waveguide); the product $dy_{n}dy'_{n}$ will be omitted in what
follows.

Averaging over $\widetilde{\gamma}_{n}$ produces:\[
\left\langle \textrm{e}^{-\mathcal{L}_{\delta}\left[\Psi\right]}\right\rangle _{\gamma}=\textrm{e}^{\sum_{n=1}^{N}\frac{x^{2}v_{n}^{2}}{8}\left\{ \int_{C_{n}}\Psi^{\dagger}\left(y_{n}\right)\varphi\left(y_{n}\right)\varphi\left(y'_{n}\right)L\,\Psi\left(y_{n}'\right)\right\} ^{2}}.\]
Then, using the Hubbard-Stratonovich transformation, we decouple the
{}``interaction terms'', introducing supersymmetric fields $Q_{n}\left(y_{n},y'_{n}\right)$
defined at each crossection $C_{n}$. To simplify the form of the
action we further assume that the leads have identical width $d$
and are attached everywhere along the perimeter of the dot. We choose
the total number $N=2\pi R/d$ ($R$ is the radius of the circle enclosing
the dot) of leads to be large, and therefore $d\ll R$. We also assume
identical transverse velocities: $v_{n}=v$. In the limit $N\rightarrow\infty$,
the density field $Q$ becomes continuous. It turns into a function
of a single variable -- the polar angle $\theta$. 

After the Gaussian integration over $\Psi\left(\mathbf{r}\right)$-variables,
an important supermatrix which needs to be determined is the effective
Green's function $\mathcal{G}\left(\mathbf{r},\mathbf{r}'\right)$.
It satisfies\[
\left(-\widehat{\mathcal{H}}_{0}-i\frac{mxva}{2}\frac{\delta\left(r-R\right)}{r}\widetilde{Q}\left(\theta\right)\right)\mathcal{G}\left(\mathbf{r},\mathbf{r}'\right)=i\delta\left(\mathbf{r}-\mathbf{r}'\right).\]
 where $\widetilde{Q}\left(\theta\right)=Q\left(\theta\right)-\left(\overline{\gamma}/2xmd\right)\Lambda$
and $a=4Rd$. The generating function takes the form \[
\left\langle Z\left[J\right]\right\rangle _{\gamma}=\int DQ\textrm{e}^{F\left[Q\right]+F_{\epsilon}\left[\mathcal{G}\right]},\]
 with the free energy\begin{align}
F\left[Q\right] & =\textrm{Str}\int d\mathbf{r}d\mathbf{r}'\left\{ -\frac{m^{2}a}{2}Q\left(\theta\right)^{2}\right.\nonumber \\
 & \left.\times\frac{\delta\left(r-R\right)}{r}\delta\left(\mathbf{r}-\mathbf{r}'\right)+\ln-i\mathcal{G}^{-1}\left(\mathbf{r},\mathbf{r}'\right)\right\} ,\label{FreeEnergy1}\end{align}
 and the symmetry breaking term $F_{\epsilon}\left[\mathcal{G}\right]=-\textrm{Str}\int\ln\left(I_{4}+\epsilon\Lambda\mathcal{G}\left(\mathbf{r},\mathbf{r}'\right)\right)d\mathbf{r}d\mathbf{r}'$.
The functional integration over $Q\left(\theta\right)$ is performed
in saddle point approximation, which requires the solution of\begin{equation}
Q_{sp}\left(\theta\right)=\frac{xv}{2m}\mathcal{G}\left(R,R,\theta,\theta,Q_{sp}\left(\theta\right)\right).\label{Saddlepoint}\end{equation}
 Assuming the solution to be diagonal and coordinate independent,
we arrive at usual structure of the saddle point: $Q_{sp}\left(\theta\right)=Q_{0}\Lambda$.
In order to analyze the fluctuations, the constant $Q_{0}$ and the
diagonal Green's function supermatrix $\mathcal{G}_{sp}\left(\mathbf{r},\mathbf{r}'\right)$
are necessary. Thus, we mapped the original problem with random boundary
condition onto the effective problem specified by the differential
equation:\begin{equation}
\left(\nabla^{2}-s^{2}\right)\mathcal{G}_{sp}\left(r,r',\theta,\theta'\right)=-\frac{2mi\delta\left(r-r'\right)\delta\left(\theta-\theta'\right)}{r},\label{ModifiedHelmholtz}\end{equation}
 and the uniform boundary condition\begin{equation}
\frac{\partial}{\partial r}\mathcal{G}_{sp}\left.\left(r,r',\theta,\theta'\right)\right|_{S^{-}}=i\frac{b\widetilde{Q}_{0}}{R}\mathcal{G}_{sp}\left.\left(r,r',\theta,\theta'\right)\right|_{S^{-}},\label{BC}\end{equation}
 where $s^{2}=-2mE$, $b=m^{2}avx$, and $S^{-}$ is the inner surface
of the dot. For both inner ($r<R$) and outer ($r>R$) domains the
solutions are readily obtained and matched together. Below we will
only need the solution for the inner domain, which can be written
in terms of the modified Bessel functions $I_{n}$ and $K_{n}$ of
$n$th order \cite{Stewartson}:\begin{align}
\mathcal{G}_{sp}\left(\mathbf{r},\mathbf{r}',s^{2}\right) & =\frac{im}{\pi}\sum_{l}I_{l}\left(sr_{<}\right)\nonumber \\
\times & \left\{ a_{l}I_{l}\left(sr_{>}\right)+K_{l}\left(sr_{>}\right)\right\} \textrm{e}^{il\left(\theta-\theta'\right)},\label{GreensfunctionIK}\end{align}
and coefficients $a_{l}$ are given via\[
a_{l}=\frac{-ibQ'_{0}K_{l}\left(sR\right)+sRK'_{l}\left(sR\right)}{ibQ'_{0}I_{l}\left(sR\right)-sRI'_{l}\left(sR\right)}.\]
 Then, $Q_{0}$ is obtained from the stationary point condition (Eq.
(\ref{Saddlepoint})). Dropping the imaginary part of $\widetilde{Q}_{0}$,
since it can be absorbed into $E$, we get\begin{equation}
\sum_{n}\frac{I_{n}^{2}\left(\widetilde{g}\right)}{\left(fI_{n}\left(\widetilde{g}\right)\right)^{2}+\left(\widetilde{g}I'_{n}\left(\widetilde{g}\right)\right)^{2}}=\frac{2\pi\left(1+\frac{\overline{\gamma}}{2xmd\widetilde{Q}_{0}}\right)}{xvb}\label{SumIdentity}\end{equation}
 Next, we set $\widetilde{g}=sR$, $f=b\widetilde{Q}_{0}$, and evaluate
the sum over $n$ in Eq. (\ref{SumIdentity}) asymptotically in the
following limit: $\widetilde{g}\gg1$, $f/\widetilde{g}\sim1$. We
replace the sum with the integral, switch to the new variable $\mu=\nu/\widetilde{g}$,
use uniform expansion for the Bessel function $I_{\nu}\left(\nu/\mu\right)$
\cite{Abramowitz} and expand the integrand in $1/\widetilde{g}$;
see Ref. \cite{Stewartson} for the details. After the substitution
$\widetilde{g}\rightarrow-ig$ ($s\rightarrow-ik$) the left-hand
side of Eq. (\ref{SumIdentity}) becomes $\pi/2\sqrt{f^{2}-g^{2}}$
to the leading order in $1/g$. Therefore, we have an algebraic equation
for $f$ (or $\widetilde{Q}_{0}$), which can be solved with the help
of the condition $k^{2}-\left(\pi/d\right)^{2}=m^{2}v^{2}$ .

At this point we introduce several relevant energy scales in {}``natural''
units $\hbar=c=1$: Thouless energy $E_{T}=k/\left(mR\right)$, mean
level spacing of a closed dot $\Delta=1/\left(mR^{2}\right)$, and
a total resonance width $\Gamma=\overline{\gamma}/\left(mdR\right)$
related to the modal decay rate of a dot having identical coupling
coefficients $\overline{\gamma}$ to all $N$ channels. In our almost
closed dot we have the following hierarchy of scales: $E_{T}\gg\Delta\gg\Gamma$.
Therefore, the set of dimensionless parameters specifying the problem
completely is given by $g=E_{T}/\Delta=kR\gg1$ (dimensionless conductance),
$\mathcal{M}=\Gamma/\Delta\ll1$ (modal overlap) together with $\overline{\gamma}$,
$x$ and $v$. Any other constants which enters Eqs. (\ref{FreeEnergy1},
\ref{Saddlepoint}), can be expressed in terms of these five. Hereafter
we continue to use the old set of parameters, including $f$ and $\widetilde{Q}_{0}$,
to keep the notation compact. The energy scales we just specified
are to be used in comparison of our predictions to experimental and
numerical results.

Now, we turn to the fluctuations around the saddle point, which can
be decomposed into a transverse piece $\delta Q^{(t)}$ (along the
saddle point manifold \cite{Efetovsbook}) and a longitudinal piece
$\delta Q^{(l)}$ (orthogonal to the saddle point manifold). The part
of the action corresponding to the $\delta Q^{(t)}$ (anticommuting
with $Q$) is given by \cite{footnote5}\begin{align*}
F_{t}\left[\delta Q\right] & =-m^{2}a\int_{0}^{2\pi}\left(\delta Q^{\left(t\right)}\left(\theta\right)\right)^{2}d\theta+\left(\frac{mxva}{2}\right)^{2}\\
 & \times\int_{0}^{2\pi}\int_{0}^{2\pi}\mathcal{G}\left(Q\right)\mathcal{G}\left(-Q\right)\delta Q^{\left(t\right)}\left(\theta\right)\delta Q^{\left(t\right)}\left(\theta'\right)d\theta d\theta'.\end{align*}
 Expanding in angular harmonics $\delta Q^{\left(t\right)}\left(\theta\right)=\sum_{l=-\infty}^{\infty}Q_{l}^{\left(t\right)}\exp\left\{ il\theta\right\} /2\pi$
and using the Ward identity (relation between product and difference
of $\mathcal{G}\left(Q\right)$ and $\mathcal{G}\left(-Q\right)$)
it is possible to show that the massive term in $F_{t}\left[\delta Q\right]$
is proportional to $mR\overline{\gamma}/\pi x\widetilde{Q}_{0}\ll1$.
Then, applying the same technique, as in the solution of Eq. (\ref{SumIdentity}),
we find that, to the leading order in $1/g$, the free energy is quadratic
in $\delta Q^{\left(t\right)}\left(\theta\right)$ for vanishingly
small $\epsilon$, and arrive at\[
F_{t}\left[\delta Q\right]\simeq-D_{0}\int_{0}^{2\pi}\left(\frac{\partial\delta Q^{\left(t\right)}}{\partial\theta}\right)^{2}d\theta,\]
 \[
D_{0}=\left(\frac{mxva\pi}{4\widetilde{Q}_{0}}\right)\frac{g\left(2g^{2}+f^{2}\right)}{f^{3}\sqrt{f^{2}-g^{2}}}.\]
 To finish the construction of the nonlinear $\sigma$-model, we integrate
out the longitudinal modes, which decouple at this order in $1/g$
from the transverse ones, and set $\delta Q^{\left(t\right)}=Q$.
Next, we expand the symmetry breaking terms $F_{\epsilon}\left[\mathcal{G}\left(Q\right)\right]$
to the lowest order in $\epsilon$. The result is given by $\left\langle Z\left[J\right]\right\rangle _{\gamma}=\int DQ\textrm{e}^{-F\left[Q\right]},$
with the free energy\begin{align}
F\left[Q\right] & =\textrm{Str}\int d\mathbf{r}d\mathbf{r}'\left\{ D_{0}\left(\frac{\partial Q}{\partial\theta}\right)^{2}\frac{\delta\left(r-R\right)}{r}\delta\left(\mathbf{r}-\mathbf{r}'\right)\right.\nonumber \\
 & \left.+\int d\theta''\epsilon\Lambda Q\left(\theta''\right)a\left(\mathbf{r},R,\theta'';\mathbf{r}',R,\theta''\right)\right\} ,\label{Sigmamodel}\end{align}
 \[
a\left(\mathbf{r},R,\theta'';\mathbf{r}',R,\theta''\right)=i\frac{mdxv}{2}\mathcal{G}_{sp}\left(R,\theta'',\mathbf{r}'\right)\mathcal{G}_{sp}\left(\mathbf{r},R,\theta''\right)\]
where $r_{>}$ ($r_{<}$) is a maximum (minimum) of $\left|\mathbf{r}\right|$
and $\left|\mathbf{r}'\right|$. The supermatrix $Q$ satisfies a
nonlinear constraint $Q^{2}=Q_{0}^{2}I_{4}$ and can be parametrized
as suggested in Refs. \cite{Efetovsbook,Mirlin} for the diffusive
case. The $n$-point correlations can be generated from the functional
given by Eq. (\ref{Sigmamodel}), which is the main result of this
paper. Just as in case of the supersymmetric nonlinear $\sigma$-model
of Ref. \cite{Efetovsbook}, the diffusion modes clearly play an important
role in the superintegrals representing correlators.

\begin{figure}[htbp]
\begin{center}\includegraphics{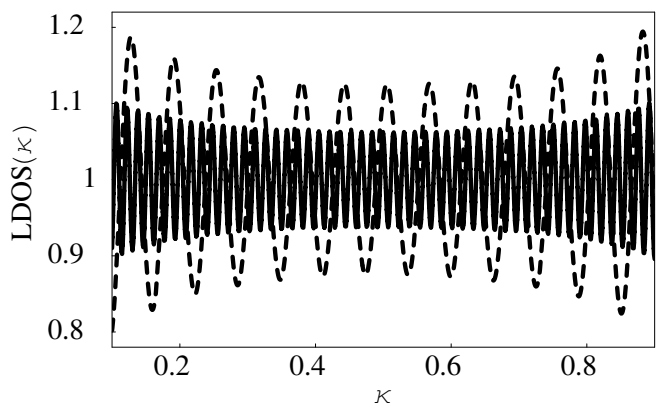}\end{center}

\caption{\label{Fig1} LDOS plotted as a function of $\kappa$ in units of
$m/\pi$ for two sets of parameters: $g=50$, $\mathcal{M}=1/12$
(dashed line); and $g=200$, $\mathcal{M}=1/3$ (solid line). In both
cases the rest of the parameters were fixed $x=0.1$; $\overline{\gamma}=0.2$;
$v=100$; }
\end{figure}

A physically measurable quantity which does not depend on fluctuations
around the saddle point is average local density of states (LDOS)
$\left\langle \rho\left(\mathbf{r}\right)\right\rangle _{\gamma}=-\left(1/\pi\right)\Im\left\langle G\left(\mathbf{r},\mathbf{r}',E\right)\right\rangle _{\gamma}.$
Our Indeed, this one-point function neither requires the knowledge
of the $\Lambda$-like structure of the saddle point manifold, nor
its explicit parametrization. It can be shown that, $\Im\left\langle G\left(\mathbf{r},\mathbf{r}',E\right)\right\rangle _{\gamma}=\left.\Im\left\langle \mathcal{G}\left(r,r',\theta,\theta',s^{2}+i\epsilon\right)\right\rangle _{Q}\right|_{s=-ik},$
where $\left\langle \dots\right\rangle _{Q}$ stands for integration
with weight $\exp\left\{ -F\left[Q\right]\right\} $ (Eq. (\ref{Sigmamodel})
). This integration reduces to evaluating the integrand at the saddle
point. Most conveniently, the average LDOS can be calculated via regularized
resolvent $\widetilde{K}\left(r,s^{2}\right)=\mathcal{G}\left(r,r,\theta,\theta,s^{2}\right)-\mathcal{G}_{0}\left(r,r,\theta,\theta,s^{2}\right)$
\cite{Stewartson}: $\pi\left\langle \rho\left(\mathbf{r}\right)\right\rangle _{\gamma}=m+\left.\Im\widetilde{K}\left(r,s^{2}\right)\right|_{s=-ik}$.
The result reads\begin{equation}
\Im\widetilde{K}\left(s^{2}\right)=\frac{m}{\pi}f\sum_{n}\frac{I_{n}^{2}\left(\kappa g\right)}{f^{2}I_{n}^{2}\left(g\right)+g^{2}\left(I'_{n}\left(g\right)\right)^{2}},\label{ImKtilda}\end{equation}
 where $\kappa=r/R$. This expression for $\Im\widetilde{K}\left(s^{2}\right)$
can be further simplified for the limiting cases: (i) $\kappa\ll1/g$,
when the main contribution to sum in Eq. (\ref{ImKtilda}) comes from
the terms with small $n$, and (ii) $\kappa\gtrsim1/g$, when we can
employ the uniform expansions of $I_{n}$. The results are summarized
in Fig. \ref{Fig1}.%

Thus, we have constructed a nonperturbative theoretical framework
to analyze one particular realization of a whole class of nanostructures:
a nearly closed system with ballistic internal dynamics interacting
randomly with the outside world through the boundary. Our approach
introduces a natural regularizer, which enables us to circumvent the
conceptual difficulties of previous approaches to closed ballistic
systems \cite{Supersymmetry}. We find that the resulting theory,
encapsulated by Eq. (\ref{Sigmamodel}), can be characterized by diffusive
modes confined to the boundary and interacting nonlocally with the
interior (see the last term of Eq. (\ref{Sigmamodel})). The supersymmetric
functional was constructed with the help of large angular momentum
modes identified as the WGM. These modes are exponentially less likely
to escape \cite{ScweitersAlfordDelos} compared to the modes with
incidence directions close to the lead normals, and consequently,
they dominate response functions at large times. Our framework should
allow us to compute the statistical properties of any physically measurable
quantity, though technical difficulties may impose strong limits.
It should be clear that our approach is also applicable, with minor
modifications, to other examples belonging to this class of systems.

Finally, the extension of this approach to generic billiards with
smooth walls (to be published elsewhere) is also possible, although
it is more technically involved \cite{arxiv}. At first sight the
non-linear supersymmetric $\sigma$-model (NLS$\sigma$M) for the
rough billiards proposed in Ref. \cite{Frahm} looks very similar
to the NLS$\sigma$Ms we derived here for open circular billiard (and
for open rough billiard in Ref. \cite{arxiv}). However, there are
several differences between two models, which can be summarized as
follows. The diffusion and (one-dimensional) localization in angular
momentum $l$ space described in Ref. \cite{Frahm} is guaranteed
by small changes in $l$ as the particle bounces off the walls. In
our case, because of the sharp edges of the region which connects
the leads to the dot the WGM trajectories may have much larger $l$
increments along the way. As a result our model describes diffusion
and localization in position (angle $\theta$) space rather than angular
momentum space.

Another issue is the role of electronic interactions. One of the possible
ways to take them into account in diffusive and ballistic systems
with large dimensionless conductance, is to use a {}``Universal Hamiltonian''
\cite{univ-ham}, which was shown to be the renormalization group
fixed point for weak interactions \cite{mm,longpaper}. We hope to
extend our analysis to the interacting ballistic case by using the
large-$N$ approach of Ref. \cite{longpaper}. We leave these questions
for future work. 

We are grateful to the NSF for partial support under DMR-0311761.


\begin{thebibliography}{10}
\bibitem{footnote3}To generate averages of products of $G^{R,A}\left(\mathbf{r},\mathbf{r}'\right)$,
which is usually the goal in any statistical analysis, one needs to
introduce source terms into action $\mathcal{L}$, such as $\int\Psi^{\dagger}\left(\mathbf{r}\right)J\left(\mathbf{r}\right)\Psi\left(\mathbf{r}\right)d\mathbf{r}$,
with $J\left(\mathbf{r}\right)$ being a coordinate dependent supermatrix,
or $\int\Psi^{\dagger}\left(\mathbf{r}\right)J\left(\mathbf{r}\right)+J^{\dagger}\left(\mathbf{r}\right)\Psi\left(\mathbf{r}\right)d\mathbf{r}$,
with $J\left(\mathbf{r}\right)$ being a coordinate dependent supervector.
These terms are not important in the derivation of the $\sigma$-model,
and we ignore them in the process.
\bibitem{footnote5}Here we omit the variation of $F_{\epsilon}\left[\mathcal{G}\right]$.
It is easily restored in the end.
\end{thebibliography}
\end{document}